\documentclass[epjST]{svjour}
\usepackage[utf8x]{inputenc}
\usepackage[pdftex,colorlinks]{hyperref}	
\usepackage{amsmath,amsfonts,amssymb,bm}
\usepackage[usenames,dvipsnames]{xcolor}
\usepackage[pdftex]{graphicx}
\usepackage{sidecap}
\usepackage{subfigure}
\usepackage{bbm}	
\usepackage{cite}	

\usepackage[nolist]{acronym}
 \acrodef{BEC}{Bose-Einstein condensate}	
 \acrodef{BZ}{Brillouin zone}			
 \acrodef{GP}{Gross-Pitaevskii} 		
 \acrodef{CL}{continuous limit} 		

\DeclareMathOperator{\diag}{diag}
\newcommand{\rmi}{{\rm i}} 

\newcommand{\vc}[1]{\bm{#1}}
\renewcommand{\r}{{\vc r}}
\renewcommand{\k}{{\vc k}}
\newcommand{\q}{{\vc q}}
\newcommand{\p}{{\vc p}}

\newcommand{\x}{{\vc x}}

\newcommand{\xpct}[1]{\bigl\langle #1 \bigr\rangle}	
\newcommand{\davg}[1]{\overline{#1}}			

\newcommand{\Ns}{\mathcal{N}}			
\newcommand{\Nc}{N_{\rm c}}			
\newcommand{\nc}{n_{\rm c}}			
\newcommand{\epn}[1]{\varepsilon^0_{#1}}	
\newcommand{\ep}[1]{\varepsilon_{#1}}		
\newcommand{\V}{{\mathcal V}}			
\newcommand{\G}{{\mathcal G}}			

\newcommand{\gh}[1]{\hat\gamma_{#1}}		
\newcommand{\ghd}[1]{\hat\gamma^\dagger_{#1}}	%

\begin{document}
\title{Bogoliubov theory on the disordered lattice}
\author{Christopher Gaul\inst{1}\fnmsep\inst{2}\fnmsep\thanks{\email{cgaul@fis.ucm.es}} \and Cord A. Müller\inst{3}}

\institute{GISC, Departamento de Física de Materiales, Universidad Complutense, E-28040 Madrid, Spain
      \and CEI Campus Moncloa, UCM-UPM, Madrid, Spain
      \and Centre for Quantum Technologies, National University of Singapore, Singapore 117543, Singapore}

\abstract{%
Quantum fluctuations of Bose-Einstein condensates trapped in disordered lattices
are studied by inhomogeneous Bogoliubov theory. Weak-disorder
perturbation theory is applied to compute the elastic scattering rate
as well as the renormalized speed of sound in lattices of arbitrary
dimensionality. Furthermore, analytical results for the condensate
depletion are presented, which are in good agreement with numerical
data.} 
 
\date{\today}

\maketitle

\section{Introduction}

In contrast to dense quantum fluids like liquid Helium, dilute gases of ultracold bosons
can be condensed into rather pure \aclp{BEC}.  
But due to interactions, there is always a fraction of non-condensed
particles, even at zero temperature
\cite{Bogoliubov1947,Penrose1956,Lee1957}. This fraction can be
enlarged by loading the atoms into the deep wells of optical lattices \cite{Xu2006}, until
finally the celebrated quantum phase transition from superfluid to
Mott insulator is reached \cite{Jaksch1998,Greiner2002,Bloch2008}. 
A somewhat different quantum phase transition from the superfluid to the
Bose-glass is driven by applying instead a
potential that varies randomly in space
\cite{Fisher1989,Gurarie2009,Sanchez-Palencia2010,Stasinska2012}. 
Generally,
the competition between kinetic energy, interaction and disorder
makes this so-called dirty boson problem very rich in its phenomenology, and
different parameter regimes require different approaches 
\cite{
Lee1990,	
Graham2009,	
Pilati2010	
}.

Here, we investigate some disorder-related properties of the deep
superfluid phase preponderant at high filling in the Bose-Hubbard model, for which 
Bogoliubov theory provides the right framework \cite{Bogoliubov1947,Oosten2001}.
Bogoliubov theory has been used extensively for disordered Bose gases \cite{Huang1992,Giorgini1994,Kobayashi2002}.
Here, we transfer our inhomogeneous Bogoliubov theory \cite{Gaul2011_bogoliubov_long}
from the continuum onto the lattice, which amounts mostly (but not
entirely) to replacing the free-space dispersion by the lattice
dispersion. The following sections \ref{model.sec}  and
\ref{secBogoliubov} describe the model and its Bogoliubov theory in
greater detail. As a first application, we compute the
effective dispersion relation with scattering rates and speed-of-sound
corrections in Sec.~\ref{effectivemedium.sec}. 

It is an essential feature of our approach that quantum fluctuations are
counted from the underlying condensate that is itself already deformed
by the external potential. In such a setting, it becomes quite a challenge to distinguish the condensed from the
non-condensed component \cite{Hu2009,White2009,Pasienski2010,Beeler2012}. Recent Quantum Monte Carlo studies have addressed
this question both for  weak lattices
\cite{Astrakharchik2011} and harmonic traps
\cite{Ray2012}. In Sec.~\ref{condep.sec}, we use the approach described in
\cite{Muller2012_momdis} to determine the disorder-induced
quantum depletion in the lattice. We reach excellent agreement with
numerical data obtained by Singh and Rokhsar \cite{Singh1994} within an explicit Bogoliubov diagonalization, 
and thus validate our analytical 
formulation. Section \ref{conclusio.sec} concludes.

\section{Bosons on a random lattice}
\label{model.sec} 
\subsection{Disordered Bose-Hubbard model} 

We consider bosonic atoms destroyed (created) by field operators $\hat
c_{\x}$ ($\hat c_\x^\dagger$) at the sites $\x = \sum_{j=1}^d n_j
\vc{a}_j$,  $n_j\in\mathbb{Z}$, of a simple Bravais lattice with basis
vectors $\vc{a}_j$, $1\le j\le d$, fulfilling the canonical commutation relations
$\bigl[\hat c_{\x}, \smash{\hat c_{\x'}^\dagger} \bigr]=\delta_{\x \x'}$,
$\bigl[\hat c_{\x}, \hat c_{\x'} \bigr]=0$, and $\bigl[\smash{\hat c_{\x}^\dagger, \hat c_{\x'}^\dagger} \bigr]=0$.
The system is governed by the grand-canonical single-band Bose-Hubbard Hamiltonian
\begin{align}
 \hat E = \sum_{\x} \biggl\lbrace
 \sum_{\x'} \hat c_{\x}^\dagger T_{\x\x'} \hat c_{\x'} 
  + (V_{\x}- \mu) \hat c_{\x}^\dagger \hat c_{\x} 
  + \frac{U}{2} \hat c_{\x}^\dagger \hat c_{\x}^\dagger \hat c_{\x}
  \hat c_{\x} 
 \biggr\rbrace\label{eqHamiltonian} .
\end{align}
The hopping matrix 
$T_{\x \x'} = -\sum_{j=1}^d J_j (\delta_{\x,\x'+\vc{a}_j} +
\delta_{\x,\x'-\vc{a}_j} - 2 \delta_{\x,\x'})$ contains  
 the tunneling amplitudes $J_j$  along
direction $j$ between nearest-neighbor Wannier
orbitals in the lowest band. To ease notations, we consider in the following  a
simple cubic lattice with edges of length
$L_j = \Ns_j a$ and a total number of sites  $\Ns = \prod_j\Ns_j$; the
generalization to other geometries 
is straightforward. The chemical
potential $\mu$ controls the particle number, and $U>0$ is the onsite
repulsion.  Equation~\eqref{eqHamiltonian} describes the 
simplest version of the disordered Bose-Hubbard model, where only the
on-site energy  $V_\x$ is random  \cite{Fisher1989,Lee1990,Singh1994}. More
generally, randomness can also occur in
hopping and interaction 
\cite{White2009,Pasienski2010}.

Without the external potential, the system is translation
invariant, so it is appropriate to use the Fourier representation 
$\hat c_{\k} =  \Ns^{-\frac{1}{2}} \sum_{\x}
 \exp\left({-i \k \cdot \x} \right)
  \hat c_{\x}$, 
with allowed wave vectors $\k$ in the first  \ac{BZ}  such that
$\sum_\k\exp[i\k \cdot (\x-\x')] = \Ns\delta_{\x\x'}$. The kinetic
energy operator becomes 
diagonal, and its matrix elements $T_{\k\k'} =\delta_{\k\k'}\epn{\k}$ 
define the clean lattice dispersion 
\begin{equation}
\epn{\k} = 2J \sum_{j=1}^d [1-\cos(a k_j)] 
\label{eqFreeDispersion} .
\end{equation}
In the \acl{CL} $a \to 0$, 
the hopping matrix turns into the continuous Laplacian, and $2J
a^2\to \hbar^2/m$, such that $\epn{\k}\to \hbar^2 k^2/2m$.
In the remainder of this contribution, we will be concerned with the inverse
mapping, namely to transfer the continuum results of 
\cite{Gaul2011_bogoliubov_long,Muller2012_momdis}  
to the lattice case.

\subsection{Statistical properties of the disorder potential}

Hitherto, cold-atom 
experimentalists have explored three ways to realize random lattices: 
(i) Sampling a continuous potential $V(\r)$, such as optical speckle,
at the lattice positions \cite{White2009,Pasienski2010}; 
(ii)  Trapping impurity atoms at or near the primary lattice sites \cite{Gavish2005,Gadway2011}; 
(iii) Creating a quasi-random distribution by adding an incommensurate
lattice to the primary lattice \cite{Roati2008}. 

The random potential $V_\x$  is characterized by its
moments $\davg{V_\x}$, $\davg{V_\x V_{\x'}}$, etc. Without loss of
generality, we can set $\davg{V_\x}=0$ by readjusting the zero of
energy. Concerning the covariance, 
one is capable of realizing, in all of the above-mentioned cases, spatially
uncorrelated disorder such that 
$ \davg{V_{\x} V_{\vc{\x'}}} =  V^2 \delta_{\x \x'}$, or 
\begin{equation}\label{VkV}
  \davg{V_{\k} V_{-\k'}} = V^2\Ns^{-1}\delta_{\k\k'}, 
\end{equation} 
for the Fourier components $V_{\k} = \Ns^{-{1}}\sum_{\x}
 \exp\left({-i \k \cdot \x} \right) V_\x$.  
In the speckle case (i), this is valid if the spatial correlation length is not
larger than  the lattice constant; for 
the Bernoulli disorder (ii) \cite{Gavish2005,Stasinska2012}, this is true by
construction, and in a quasi-random system (iii)
this requires the incommensurability to be large enough so that 
no repetitions occur in the finite size under study \cite{Modugno2009}.

\subsection{Condensate deformation on the mean-field level}
\label{condef.sec}
\begin{figure}[tb]
\centerline{\includegraphics[width=0.85\linewidth]{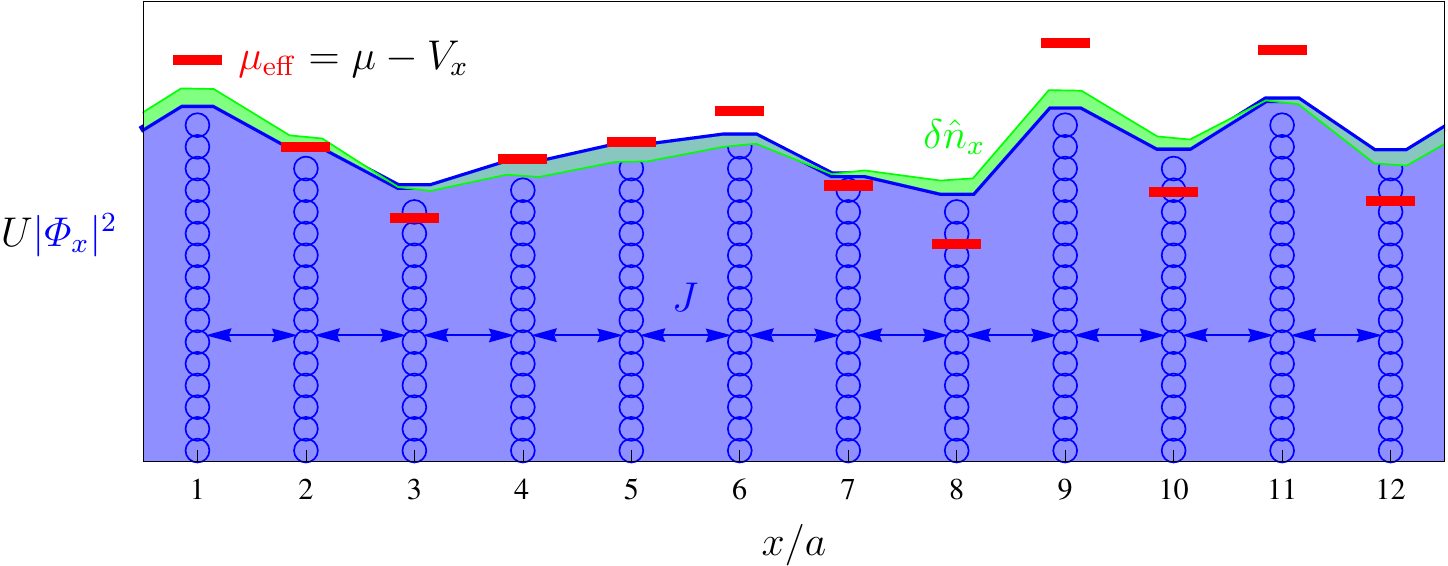}}
\caption{Schematic view of a disordered lattice hosting an extended, but deformed condensate.
In the atomic limit $J=0$, the condensate density (blue) would fill the
wells up to $(\mu-V_\x)/U$. Finite hopping
$J$ (in this plot $J=U\nc$) couples adjacent sites and smoothens the profile. 
Section~\ref{secBogoliubov} and following discuss the quantum fluctuations 
$\delta \hat n$ (and $\delta\hat\varphi$), here cartooned in green.}
\label{figSketchDisorder} 
\end{figure}

When the discreteness of the 
particle number at a given site is negligible, i.e., for high particle
numbers per site, $n\gg U/J$, a mean-field description for 
the extended condensate is a good starting point (see the schematic representation in Fig.\ \ref{figSketchDisorder}). Replacing the
operators $\hat c_\x^{(\dagger)}$ by their ground-state expectation
value $\Phi_\x$ (which can be chosen real) and minimizing \eqref{eqHamiltonian} yields the
\acl{GP} equation, which we write in the Fourier representation: 
\begin{align}
 \mu \phi_{\k} = \epn{\k} \phi_{\k} 
     + \sum_{\k'} V_{\k-\k'} \phi_{\k'}
     + \frac{U}{\Ns}\sum_{\k' \k''} \phi_{\k-\k'} \phi_{\k'-\k''}
     \phi_{\k''}.  \label{eqGPE}
\end{align}
If the external potential is weak, the
 homogeneous solution $\phi_{\k}^{(0)} = \Nc^{1/2} \delta_{\k 0}$ is only
 slightly perturbed, and one can compute 
the deformed condensate amplitude   $\phi_\k =
\phi_{\k}^{(0)} +\phi_{\k}^{(1)} + \dots$ perturbatively in powers of
$V$  \cite{Lee1990}. At fixed number of condensed particles, $\sum_{\k}
|\phi_\k|^2=\Nc$, also the chemical
potential has to be expanded. 
As a result, the condensate deformation reads  $\phi_\k^{(1)} =
-\Nc^{1/2} \tilde V_\k$ to first order in the reduced potential matrix element  
\begin{align}
  \tilde V_\k = \frac{1-\delta_{\k 0}}{\epn{\k}+2U\nc} V_{\k}.  
  \label{tildeVk}
\end{align}
This expression matches Eq.~(9) in
\cite{Gaul2011_bogoliubov_long} [or Eq.~(15) in \cite{Lee1990}], after substituting $gn$ by $U\nc=U
\Nc/\Ns$, and reading $\epn{\k}$ as the lattice dispersion
\eqref{eqFreeDispersion}. 
Also all higher-order terms agree with
their respective expressions in \cite{Gaul2011_bogoliubov_long}, as do derived
quantities like the inverse condensate amplitude and densities. 

In the limit $J\ll U \nc$ of weak coupling between sites (which requires large
occupation such that still $J\nc\gg U$), the bare kinetic energy
$\epn{\k}$ in the denominator of the reduced potential \eqref{tildeVk}
is always negligible compared to $2U\nc$. Thus, the condensate  
faithfully follows the potential, to all orders, resulting in a Thomas-Fermi profile  
and a screened disorder landscape \cite{Lee1990}. In the strong-coupling limit $J\gg U \nc$, kinetic and interaction energy
become comparable at a characteristic wave vector equal to 
the inverse of the healing length, $\xi = a \sqrt{J/(U\nc)}\gg a$. Whereas potential fluctuations on large scales
$k^{-1}\gg\xi$ are still followed by the
condensate, fluctuations on small scales $k^{-1}\ll\xi$ cost
too much kinetic energy, and result only in a smoothed imprint \cite{Sanchez-Palencia2006}. 

The deformed condensate consists of particles with the
(ensemble-averaged) momentum distribution \cite{Muller2012_momdis}
\begin{equation}\label{eqCondensateMomdis}
n_{\text{c}\k} = \davg{|\phi_\k^2|} = \Nc\left[ (1-V_2)\delta_{\k0} +
  \davg{|\tilde V_\k^2|}\right];
\end{equation}  
the second equality holds to order $V^2$ in the disorder strength. 
Here, $V_2=\sum_\k \davg{|\tilde V_\k^2|}$ is the 
fraction of condensed particles with $\k\neq0$, sometimes referred to as glassy fraction or Edwards-Anderson order parameter \cite{Yukalov2009}. In the
weak-coupling limit $J\ll U \nc$ and for uncorrelated disorder \eqref{VkV}, this
fraction is $V_2 = v^2/4$, where $v^2=(V/U\nc)^2 $ is the 
disorder variance in units of the chemical potential of the homogeneous lattice gas.  

\section{Inhomogeneous Bogoliubov Hamiltonian on the lattice}\label{secBogoliubov}

In this section, we set up the inhomogeneous Bogoliubov Hamiltonian
for quantum fluctuations $\delta \hat c_{\x}$ around the
(deformed) mean-field condensate $\Phi_\x$ on a disordered lattice. 
For the reasons given in \cite{Gaul2011_bogoliubov_long}, a well-defined 
theory ensues if one expresses the fluctuations via 
density and phase, 
\begin{align}
 \delta \hat c_{\x} &= \hat c_\x - \Phi_\x = 
\frac{\delta \hat n_{\x}}{2 \Phi_{\x}} + \rmi \Phi_{\x} \delta \hat
\varphi_{\x} \label{eqDensPhase}. 
\end{align}
Inserting \eqref{eqDensPhase} into
\eqref{eqHamiltonian}, and collecting all second-order 
terms,%
\footnote{First-order terms vanish because of \eqref{eqGPE}. Higher-order terms describe interactions among the
excitations that remain negligible for very low temperatures and dilute
gases.}  
one arrives at a fluctuation Hamiltonian of the form
\begin{equation}\label{eqHBogoliubovSR}
 \hat H = \frac{1}{2}\sum_{\k \k'} 
   \left[
    \nc \delta\hat\varphi_{\k}^\dagger S_{\k \k'} \delta \hat\varphi_{\k'}
  + \frac{1}{4\nc}\delta\hat n_{\k}^\dagger R_{\k \k'} \delta \hat n_{\k'}
   \right] .
\end{equation}
Crossed terms have canceled, up to constants from commutators, and
the coupling matrices between different Fourier modes are found to be 
\begin{align} 
 S_{\k \k'} & = \frac{2}{\Nc}\sum_\p
 \phi_{\k-\p}\phi_{\p-\k'}(\epn{\p}-\epn{\k-\p}), \label{Skkpr}\\
R_{\k \k'} & =  \frac{2}{\Nc}\sum_{\p,\q}
\check\phi_{\k-\p}\check\phi_{\p-\k'-\q}\left[(\epn{\p} -\mu)\delta_{\q0}+
  V_\q \right] + 6 U\nc\delta_{\k\k'}. \label{Rkkpr}
\end{align} 
Here, $\check\phi_\k = [\nc/\phi]_\k$ are Fourier components of the
inverse condensate amplitude. 
Eqs.~\eqref{Skkpr} and \eqref{Rkkpr} are the lattice equivalents of (29) and (30) in
\cite{Gaul2011_bogoliubov_long}. There, we used the
identity $\check\phi\nabla \phi = - \phi\nabla \check\phi$ together
with \eqref{eqGPE}
to express $R_{\k\k'}$ as a quadratic functional of
$\check\phi_\k$ without appearance of $V_\q$ nor $\mu$, which turns out
impossible on the lattice.

\subsection{Homogeneous case}

\begin{figure}[tb]
 \subfigure[]{\raisebox{0.2ex}{\includegraphics[width=0.333\linewidth]{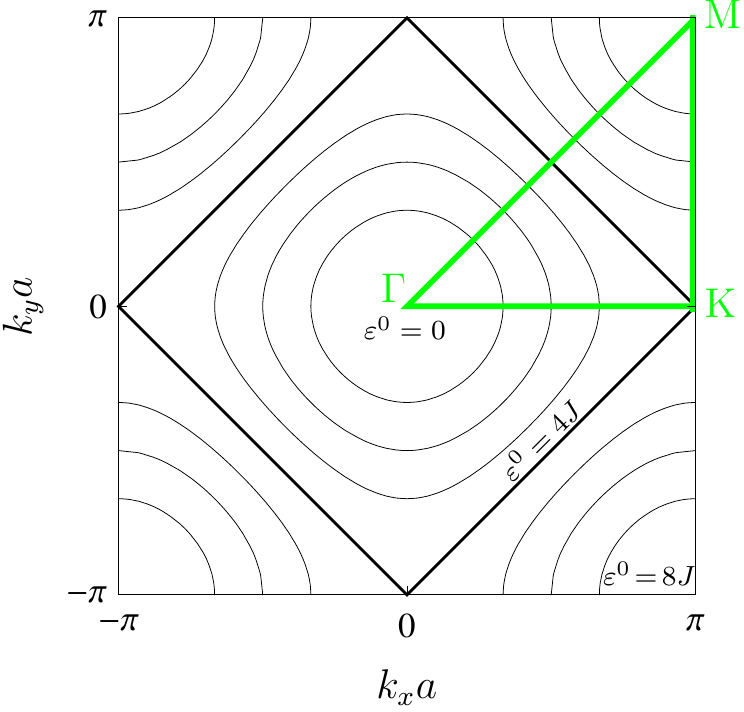}}\label{figBZ}}
 \subfigure[]{\includegraphics[width=0.625\linewidth]{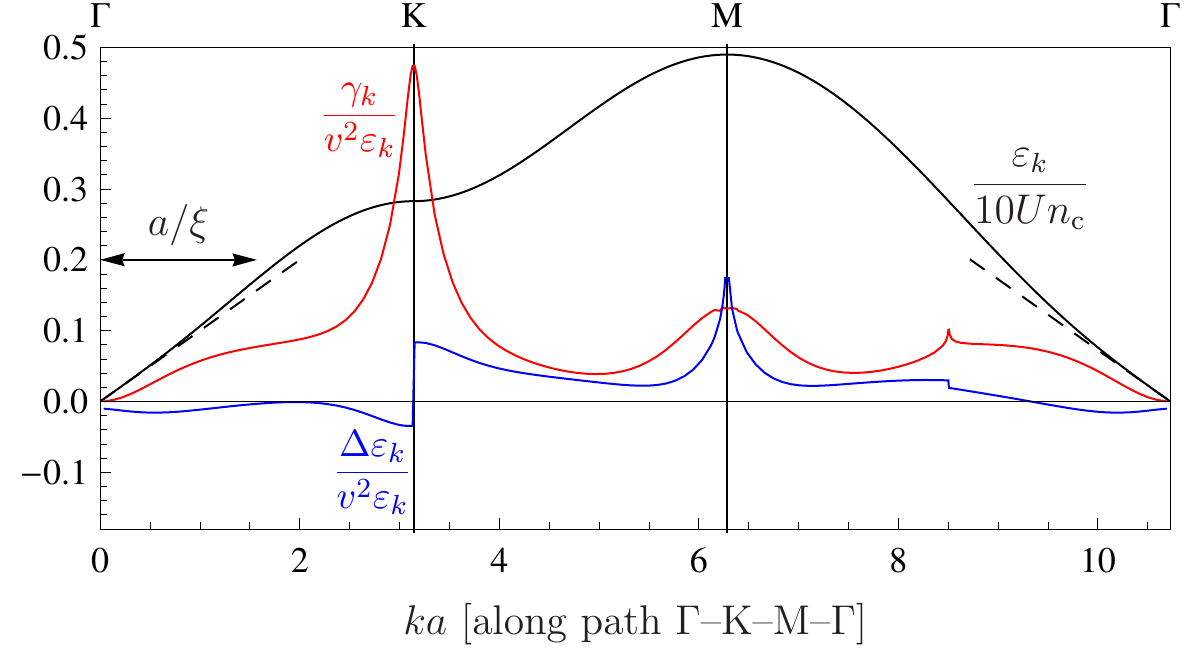}\label{figEffectiveMediumCorrections}}
 \caption{(a) Equipotential lines of the clean and Bogoliubov
   lattice dispersions, \eqref{eqFreeDispersion} and
   \eqref{eqHomBgHam}, respectively, in the first \ac{BZ} of the 2D lattice. 
  (b) Plots  of several quantities for $J=0.5U\nc$ along the closed path connecting $\mathrm{\Gamma}$,
  K, and M in (a):  
  Clean Bogoliubov dispersion $\ep{\k}$, Eq.~\eqref{eqHomBgHam},
  in units of $10U\nc$ (full black line), together with 
the phonon dispersion $\ep{\k}=\hbar c k$ for $k\xi\ll1$ (dashed black). 
 Colored curves show the disorder corrections calculated in
 Sec.~\ref{effectivemedium.sec}:  
 relative scattering rate $\gamma_\k/\ep{\k}$ (red);
 relative dispersion shift 
 $\Delta\ep{\k}/\ep{\k}$ (blue), both in units of
 $v^2=V^2/(U\nc)^2$.}
\end{figure}

In absence of the potential $V_\x$, the condensate is homogeneous,  
and the coupling matrices reduce to $S_{\k\k'}^{(0)} =
2\epn{\k}\delta_{\k\k'}$ and $R_{\k\k'}^{(0)} = 
2[\epn{\k}+2U\nc]\delta_{\k\k'}$. The Hamiltonian
becomes diagonal, $ \hat H^{(0)} 
  = \sum_\k \ep{\k} \ghd{\k} \gh{\k}$, 
after a canonical transformation
to Bogoliubov quasiparticles (``bogolons'') such that 
$[\gamma_{\k'},\gamma^\dagger_\k]=\delta_{\k\k'}$. This is achieved by
\begin{equation} \label{bgtrafo}
\begin{pmatrix}
\gh{\k} \\
\ghd{-\k}
\end{pmatrix} 
= A_\k \begin{pmatrix}
 \rmi \sqrt{\nc}\delta\hat\varphi_\k  \\
\delta \hat n_\k /(2\sqrt{\nc}) 
\end{pmatrix},
\qquad 
A_\k=  
\begin{pmatrix}
 a_\k & a_\k^{-1}\\
-a_\k & a_\k^{-1}
\end{pmatrix} 
\end{equation}  
with 
parameter $a_\k = (\epn{\k}/\ep{\k})^{1/2}$ in each
$\k$-sector. The resulting Bogoliubov dispersion 
\begin{equation}\label{eqHomBgHam}
 \ep{\k} =
  [\epn{\k}(\epn{\k} + 2 U \nc)]^{1/2}
\end{equation}
is plotted in Fig.~\ref{figEffectiveMediumCorrections} along a
representative path in the  \ac{BZ} of a 2D simple
cubic lattice shown in Fig.\ \ref{figBZ}. For the chosen parameters, the healing
length 
$\xi= 2 a/\pi $ separates the sound-wave regime $k\xi \ll1$ with linear
dispersion from the particle-like excitation regime 
$k\xi\gg1$ with parabolic dispersion.

\subsection{Bogoliubov scattering vertex}

For any given realization of the external potential $V_\x$, and
thus a given realization of the coupling matrices \eqref{Skkpr} and
\eqref{Rkkpr}, one can diagonalize the quadratic Hamiltonian
\eqref{eqHBogoliubovSR}. 
Analytical calculations for weak disorder, however, are best done in the 
bogolon basis \eqref{bgtrafo}---with
density (and phase) fluctuations counted from the deformed condensate. The price
for keeping a plane-wave basis is the 
appearance of an impurity-scattering term in the
Hamiltonian, which can be treated using standard perturbation theory. 

We separate the homogeneous contributions
from the coupling matrices by defining $R_{\k \k'}^{(V)} = R_{\k \k'}-R_{\k \k'}^{(0)}$, as well
as $S_{\k \k'}^{(V)} =S_{\k \k'}-S_{\k \k'}^{(0)}$. By the canonical
transformation \eqref{bgtrafo} to Nambu pseudo-spinors $\hat
\Gamma_\k^\dagger = (\ghd{\k},\gh{-\k})/\sqrt{2}$,
the fluctuation Hamiltonian \eqref{eqHBogoliubovSR} is brought into
impurity-scattering form \cite{Gaul2011_bogoliubov_long}:
\begin{align}
 \hat H = \sum_\k \ep{\k} \hat \Gamma_{\k}^\dagger \hat{\Gamma}_{\k}
  + \sum_{\k, \k'}
      \hat\Gamma_\k^\dagger
       \V_{\k \k'}       \hat\Gamma_{\k'}. \label{eqHamiltonianBogolon}
\end{align}
The bogolon scattering vertex%
\footnote{Eq.~(31) in \cite{Gaul2011_bogoliubov_long} contains a
  misprint with two factors of 1/2 too much; the following expressions
are not affected.}
$\V_{\k \k'} = (A_{\k}^{-1})^\dagger \diag\bigl(S^{(V)}_{\k \k'},R^{(V)}_{\k
  \k'}\bigr) A_{\k'}^{-1}$ 
is a function of the condensate ground-state configuration via \eqref{Skkpr}
and \eqref{Rkkpr}, and  
can be expanded in powers of the bare potential using the perturbation
theory of Sec.~\ref{condef.sec}. Thus, our approach goes beyond earlier 
Bogoliubov theories on disordered lattices \cite{Lee1990,Singh1994} 
inasmuch our Hamiltonian is useful even for purely analytical
calculations. To first
order, 
\begin{align}
 R_{\k \k'}^{(1)} &= 2 {\tilde V_{\k-\k'}} \left[\epn{\k-\k'} + \epn{\k} + \epn{\k'}\right], \label{eqR1kks}\\
 S_{\k \k'}^{(1)} &= 2 {\tilde V_{\k-\k'}} \left[\epn{\k-\k'}-
   \epn{\k} - \epn{\k'}\right]. \label{eqS1kks}
\end{align} 
If one is interested in results to second order at most, at this order only disorder-averaged,
diagonal terms are needed:  
\begin{align}  
 \davg{S_{\k \k'}^{(2)}} 
  &=2  \delta_{\k\k'}  \sum_{\q} \davg{\bigl|\tilde V_{\q}\bigr|^2}  \left[ \epn{\k-\q} - \epn{\k}-\epn{\q} \right] ,
\label{eqAvgS2kks} \\
   \davg{R_{\k \k'}^{(2)}} &=2  \delta_{\k\k'}  \sum_{\q}  \davg{\bigl|\tilde
     V_{\q}\bigr|^2}  \left[\epn{\k-\q}  +3\epn{\k} +3\epn{\q} \right]   . \label{eqAvgR2kks}
\end{align}
Contrary to the continuous-limit case \cite{Gaul2011_bogoliubov_long}, where $S_{\k\k'}$ of Eq.\ (28) is
proportional to $n_{\text{c}\k-\k'}$, here its diagonal elements have no reason to vanish.

\section{Effective medium theory}
\label{effectivemedium.sec}

One of the first things one may want to know is how the disorder
modifies the excitation spectrum. Clearly, randomness limits the
life-time of plane-wave excitations, affects the speed of sound,
changes the density of states, and causes localization
\cite{Kuhn2007a,Gaul2009a,Lugan2007a,Lugan2011,Gaul2011_bogoliubov_long}. 
All these effects can be assessed by applying  
diagrammatic perturbation theory to the Hamiltonian
\eqref{eqHamiltonianBogolon}, for instance via a Green's function
formalism. 
The Nambu-Green function of the clean Hamiltonian  is 
$\G_{0\k}(z) = \diag\bigl(G_{0\k}(z),G_{0\k}(-z)\bigr)$, with 
$G_{0\k}(z)=[z - \ep{\k}]^{-1}$, and the full
Green function obeys the recursive equation $\G = \G_0 + \G_0\V\G$.  
The ensemble-averaged Green function $\davg{\G} = \G_0 + \G_0 \Sigma
\davg{\G}$ describes the propagation of
bogolons through an effective medium, whose effect is to renormalize the 
dispersion:
\begin{equation}\label{dispersion} 
\hbar \omega_\k 
 = \ep{\k} + \Sigma_{11} (\hbar\omega, \k) = 
 \ep{\k} + \Delta \ep{\k} - \rmi
\gamma_\k/2 .
\end{equation}
$\Sigma_{11}$ designates the first block of the Nambu self-energy. 
To second order in the bare potential,  
one finds two types
of terms, $\Sigma^{(2)} = \davg{\V^{(1)}\G_0\V^{(1)}} +
\davg{\V^{(2)}}$, 
resulting from the perturbative expansion of the bogolon vertex. 

Let us first discuss the elastic scattering rate  $\gamma_\k$. 
To lowest order, known as the Born approximation,   
it is given by Fermi's Golden Rule
via a momentum integral over the energy shell:  
\begin{equation}
\gamma_\k =  \frac{\pi}{8} \sum_\p \davg{\bigl|a_\k a_{\p}
  R_{\k\p}^{(1)} + a_\k^{-1}a_{\p}^{-1}S^{(1)}_{\k\p} \bigr|^2}
\delta(\ep{\k}-\ep{\p}).  
\end{equation}
Due to the rather complicated form of the energy shells, even for a
simple cubic lattice, the integral has to be evaluated
numerically.%
\footnote{We parametrize the integral by the polar (and azimuthal, in
  $d=3$) angle, determine the modulus of $\p$ such that $\ep{\p}=\ep{\k}$, 
 and evaluate the integrand there.
This has to be multiplied with the surface element and the Jacobian $[\partial \ep{\p}/\partial
p_n]^{-1}$, where $p_n$ is the component normal to the energy shell. 
For high energies, we take the corner M as origin.} 
In Fig.~\ref{figEffectiveMediumCorrections}, the relative scattering
rate $\gamma_\k/\ep{\k}$ in units of $v^2$ is shown across the \ac{BZ}. Just as in the
\acl{CL}, the scattering of low-energy excitations near the  $\mathrm{\Gamma}$
point is suppressed, resulting in long-lived, soft modes. Strong scattering
occurs near the symmetry point K at the \ac{BZ} boundary. 

The real part of the self-energy in \eqref{dispersion} shifts the energy of excitations.
In Fig.~\ref{figEffectiveMediumCorrections}, the shift $\Delta \ep{\k}/\ep{\k}$ relative to the clean dispersion is plotted in units of
$v^2$.
The shift $\Delta \ep{\k}$ is a principle-value integral over a kernel with a simple pole at the energy shell \cite{Gaul2011_bogoliubov_long}.
As shown in Fig.\ \ref{figBZ}, the energy shell changes its topology at $\varepsilon = [4J(4J+U\nc)]^{1/2}$, which causes jumps in the shift of the dispersion $\Delta \ep{\k}$.
For the soft modes, $\Delta\varepsilon_\k$ results in a changed speed of sound, $\davg{c} = c + \Delta \davg{c}$, which we observe to be negative in the 2D example of Fig.\ \ref{figEffectiveMediumCorrections}.
In the limit $ J/U\nc = \xi^2/a^2 \gg 1$, the lattice constant $a$ drops out, and we recover the asymptotics for uncorrelated disorder in the continuum,
$\Delta\davg{c}/c \propto v_\delta^2 = v^2 (a/\xi)^d$, where the proportionality factor is negative in 1D, positive in 3D and vanishes in 2D \cite{Gaul2011_bogoliubov_long}. 
The increase in 3D, 
$\Delta\davg{c}/c = 5   v_\delta^2 / (48\sqrt{2}\pi)$, coincides with the established result \cite{Giorgini1994,Lopatin2002,Falco2007} for uncorrelated disorder. 
However, as noted  in \cite{Gaul2011_bogoliubov_long}, an increase of $c$ with disorder is untypical.
Generically, one expects the disorder to reduce the speed of sound, and consequently to increase the
low-energy density of states.
And really, in the opposite, weak-coupling limit  $ J \ll U\nc$,
the shift is negative and reduces to the value 
$\Delta\davg{c}/c = -0.375v^2, -0.215v^2, -0.177v^2$ in dimensions 1, 2, and 3, respectively.
This is very similar to the hydrodynamic regime of the continuum \cite{Gaul2011_bogoliubov_long,Gaul2009a}, but with slightly different constants.

\section{Condensate depletion}
\label{condep.sec} 

\begin{figure}[tb]
 \subfigure[]{\raisebox{1ex}{\includegraphics[width=0.425\linewidth]{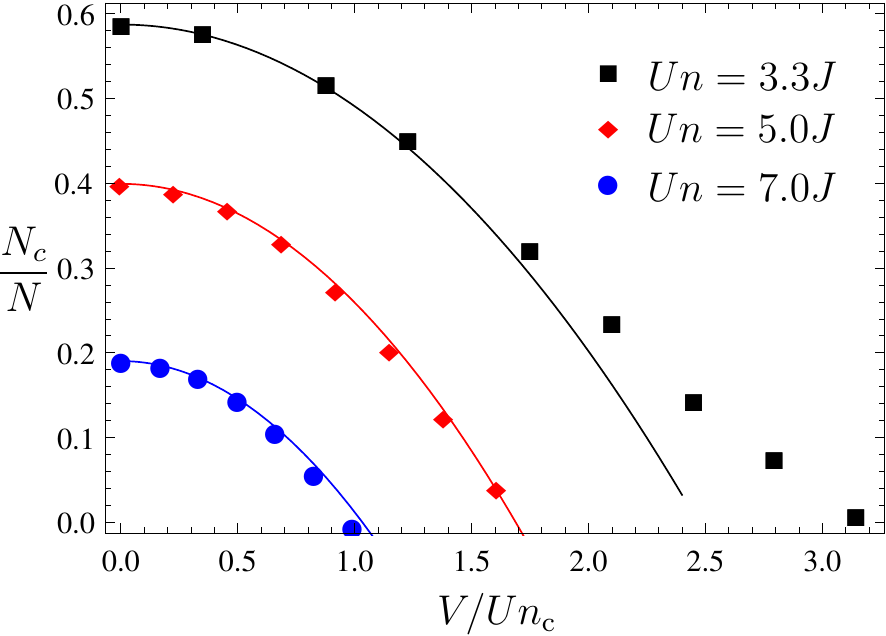}\label{figCondepSR}}}\hfill
 \subfigure[]{\includegraphics[width=0.545\linewidth]{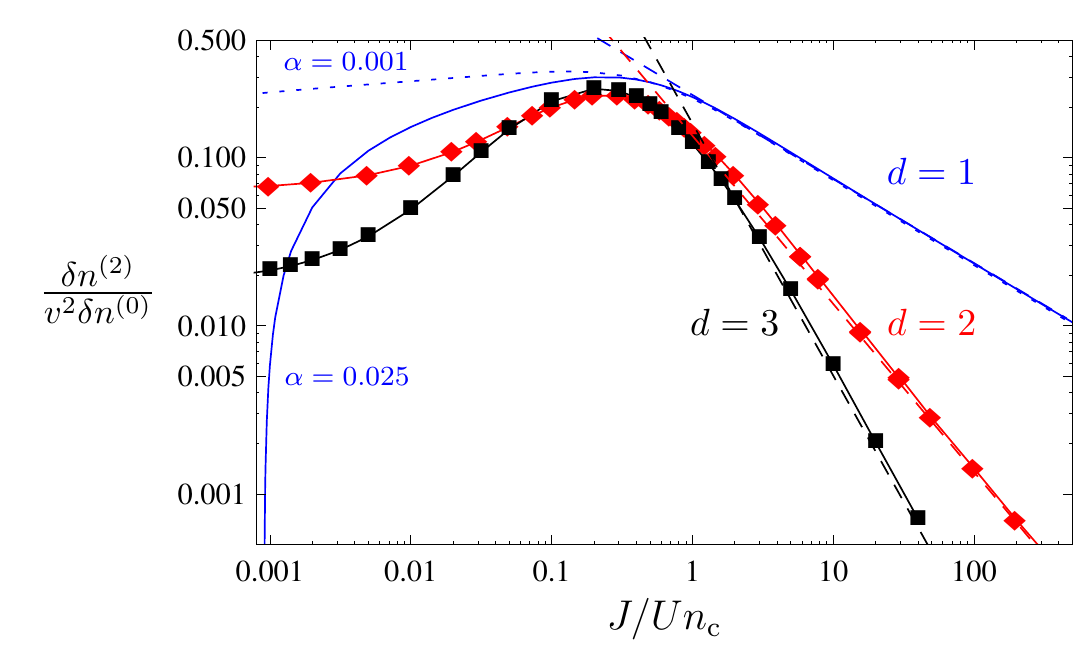}\label{figCondep}}
 \caption{(a) 2D condensate fraction $\Nc/N$ as function of the disorder strength $v=V/U\nc$ for different values of $U\nc/J$. 
 The data points are taken from Fig.~4 in \cite{Singh1994}, the result of an exact Bogoliubov diagonalization.
 The lines show our results for the condensate fraction resulting from the combined clean and 
 potential depletion $\delta n = \delta n^{(0)} + \delta n^{(2)}$, Eqs.~\eqref{deltan0} and \eqref{deltan2}.
 For not too strong disorder $v=V/U\nc \lesssim 1.5$ the agreement is excellent.
 (b) Disorder-induced depletion, relative to the clean depletion, in dimensions $d=1,2,3$.
  The dashed lines show the large-$J$ asymptotics $\delta n^{(2)}/\delta n^{(0)} \approx \beta_d v_\delta^2 = v^2 (U\nc/J)^{d/2}$, corresponding to the uncorrelated limit of Ref.~\cite{Muller2012_momdis}.
  In one dimension, an infrared cutoff $\alpha = k_{\rm IR} \xi$ is needed.}
\end{figure}

Due to the repulsive interaction between atoms, there is always a certain 
fraction of particles outside of the condensate. In a weakly interacting gas, 
this fraction can be estimated using Bogoliubov theory. 
Let us first discuss the homogeneous lattice case. Condensation occurs in the state with 
quasimomentum $\k=0$, and the non-condensed particles have a momentum distribution 
$\delta n_\k^{(0)} = \xpct{\delta \hat c_\k^\dagger \delta \hat c_\k} =
(a_\k-a_\k^{-1})^2/4$, as follows from \eqref{eqDensPhase} and
\eqref{bgtrafo} in the Bogoliubov vacuum at $T=0$. The total density
of non-condensed particles (number per lattice site) is  
\begin{equation}\label{deltan0}
 \delta n^{(0)}  = \Ns^{-1} \sum_\k \frac {(\ep{\k}-\epn{\k})^2} {4\ep{\k}\epn{\k}}.
\end{equation}
In the \acl{CL}, $J \gg U\nc$, it shows the well-known scaling 
proportional to $(U\nc/J)^{d/2} = (a/\xi)^d$.
In the weak coupling regime, $J\ll U \nc$, the asymptotics is 
$\delta n^{(0)} \approx g_d \sqrt{U\nc/J}$, where $g_3=0.161$ and $g_2=0.227$.
In one dimension, an infrared cutoff is needed, because there is no
true Bose-Einstein condensation, and $\delta n^{(0)}$ grows
logarithmically with the cutoff. 
In all dimensions, the condensate fraction, or relative
number of particles in the condensate, is $\Nc(0)/N = 1-\delta
n^{(0)}/n$. Using \eqref{deltan0} in the thermodynamic limit, we find
excellent agreement with the clean condensate fraction shown in Fig.~4
of Singh and Rokhsar \cite{Singh1994}, and infer $n=0.33$ particles
per site for their data. Even though the density is quite low, and the
system parameters do not verify the prima facie condition
$U\ll J n$ of validity
for mean-field theory, at this fractional occupation, the clean system has a
superfluid ground state. Nonetheless, we cannot expect Bogoliubov
theory to yield accurate predictions as the condensate fraction falls
below 50\% \cite{Oosten2001,Xu2006}. 

Next, we study the excitation momentum distribution  in the presence of disorder. 
We can express $\delta n_\k$, still via \eqref{eqDensPhase} and
\eqref{bgtrafo}, in terms of the condensate profile $\Phi_\x$ and
bogolon ground-state expectation values such as 
$\xpct{\hat \gamma_\k^\dagger \hat \gamma_{\k'}}$;
for details see \cite{Muller2012_momdis}.
To second order in the disorder potential, the result can be written   
\begin{equation}\label{deltan2} 
 \delta n_\k^{(2)} = \sum_{\p} \tilde M_{\k \p}^{(2)} |\tilde V_{\k-\p}|^2. 
\end{equation}
Now the kernel $\tilde M_{\k \p}^{(2)}$ of Eq.~(49) in
\cite{Muller2012_momdis} has to be evaluated using the lattice
dispersion \eqref{eqFreeDispersion} and the envelope functions that
follow from  
\eqref{Skkpr} and \eqref{Rkkpr}.   
Just as in the continuous case of \cite{Muller2012_momdis}, the
momentum distribution of excitations induced by the disorder broadens
the condensate momentum distribution, Eq.~\eqref{eqCondensateMomdis}. 

The resulting, additional quantum depletion of the condensate density
caused by the disorder potential, or potential depletion for short, is
the sum over the corresponding momentum distribution, 
$\delta n^{(2)} = \Ns^{-1} \sum_\k \delta n^{(2)}_\k$.
In Fig.\ \ref{figCondepSR}, we show results for two dimensions and
compare them to data from the exact Bogoliubov diagonalization of
Ref.\ \cite{Singh1994}. 
Within our perturbation theory, we can only estimate the quadratic
corrections of order $V^2$, but this appears to be a rather good
approximation even for strong disorder up to $V\approx U\nc$.  
The good agreement validates our analytical approach to inhomogeneous
Bogoliubov theory. 

In contrast to \cite{Singh1994}, our theory is analytical up to the
evaluation of the integrals, and we do not need to average over
realizations of disorder. 
This allows us to consider higher dimensions with little additional effort.
In Fig.~\ref{figCondep}, the relative potential depletion is shown as function of $J/U\nc$ in dimensions $d=1,\,2,\,3$.
In all cases, the depletion takes its maximum around $J \approx 0.5 U\nc$.
The large-$J$ behavior coincides with the limit $\sigma \to 0$ of the continuous case.
In accordance with \cite{Muller2012_momdis}, we find $\delta
n^{(2)}/\delta n^{(0)} \approx \beta_d v_\delta^2$, with
$\beta_1 \approx 0.236$ (slightly cutoff-dependent), $\beta_2 \approx 0.135$, and
$\beta_3 \approx 0.160$.  

In the weak coupling regime, $J \ll U\nc$, $\delta n^{(2)}$ shows the same scaling proportional to $\sqrt{U\nc/J}$ as $\delta n^{(0)}$, such that the ratio $\delta n^{(2)}/v^2 \delta n^{(0)}$ tends to a constant ($\approx 0.016$ in 3D, and $\approx 0.06$ in 2D).
In one dimension, however, the weak-coupling behavior is
cutoff-dependent: as $J/U\nc$ gets smaller than the cutoff $\alpha$,
the ratio diminishes and it is not possible to obtain a well-defined
limit, as becomes evident from the two curves with different cutoffs shown in Fig.\
\ref{figCondep}. 

To summarize this section,  we have shown that the disorder potential does indeed
increase the condensate depletion, essentially because it deforms the
condensate and creates high-density regions, where the
interaction-induced depletion is enhanced. The total depletion can be
written \cite{Muller2012_momdis}
\begin{equation} 
\delta n= \delta n^{(0)} \left [1+v^2 \Delta (J/U\nc) \right],
\end{equation} 
where $\Delta(J/U\nc)$, plotted in \ref{figCondep}, is always smaller
than unity. Thus, the potential depletion for not-too-strong disorder
remains a relatively small correction, comparable to the homogeneous
depletion, up to $V\sim U\nc$ at least. For a large mismatch between
$J$ and $U\nc$, the condensate becomes even more resilient to the
disorder by virtue of smoothing (for $J\gg U\nc$) and screening (for
$J\ll U\nc$).

\section{Conclusions}
\label{conclusio.sec} 

We have studied the superfluid phase of the disordered Bose-Hubbard
model by Bogoliubov theory, transferring our previous continuum formulation to
the lattice case. The key point is to describe the quantum
fluctuations around a condensate that is deformed by the random
potential. This approach permits purely analytical
calculations, at least to second order in the disorder strength. We
have determined the renormalized dispersion relation, and find strong scattering at the K
point (3D: X and M points), where the free dispersion relation has a
saddle point. Otherwise, the excitations stay gapless, and soft modes
are long lived, with well-defined corrections to the speed of sound. 
Furthermore, we have calculated by how much 
the non-condensed fraction increases due to the disorder. This
additional depletion is found to agree with earlier, numerical results
by Singh and Rokhsar. We determine its dependence on the system
parameters and find that the disorder-induced depletion relative to
the clean depletion is greatest when the hopping matches the chemical
potential, which should provide useful information for further experimental
studies.

\begin{acknowledgement}
Research of C.G.\ was supported by a PICATA postdoctoral fellowship
from the Moncloa Campus of International Excellence (UCM-UPM). 
C.M.\ acknowledges financial support from Fondation des Sciences
Math\'ematiques de Paris (FSMP) within the programme ``Disordered
Quantum Systems'' at Institut Henri Poincar\'e. CQT is a Research Centre of Excellence funded by the Ministry of 
Education and the National Research Foundation of Singapore. 
C.G.\ thanks Jürgen Schiefele and Martin Heimsoth for helpful discussions.
\end{acknowledgement}

\bibliographystyle{mybst_noeprint_notitle_doi-link}
\bibliography{references_bogolattice}

\begin{thebibliography}{10}
\newcommand{\enquote}[1]{``#1''}
\newcommand{\Doi}[2]{\href{http://dx.doi.org/#1}{#2}}

\bibitem{Bogoliubov1947}
N.~Bogoliubov, Journal of Physics (Moscow) \textbf{11}, 23 (1947).

\bibitem{Penrose1956}
O.~Penrose and L.~Onsager, \Doi{10.1103/PhysRev.104.576}{Phys. Rev.}
  \textbf{104}, 576 (1956).

\bibitem{Lee1957}
T.~D. Lee, K.~Huang, and C.~N. Yang, \Doi{10.1103/PhysRev.106.1135}{Phys. Rev.}
  \textbf{106}, 1135 (1957).

\bibitem{Xu2006}
K.~Xu, Y.~Liu, D.~E. Miller, J.~K. Chin, W.~Setiawan, and W.~Ketterle,
  \Doi{10.1103/PhysRevLett.96.180405}{Phys. Rev. Lett.} \textbf{96}, 180405
  (2006).

\bibitem{Jaksch1998}
D.~Jaksch, C.~Bruder, J.~I. Cirac, C.~W. Gardiner, and P.~Zoller,
  \Doi{10.1103/PhysRevLett.81.3108}{Phys. Rev. Lett.} \textbf{81}, 3108 (1998).

\bibitem{Greiner2002}
M.~Greiner, O.~Mandel, T.~Esslinger, T.~W. H{\"a}nsch, and I.~Bloch,
  \Doi{10.1038/415039a}{Nature} \textbf{415}, 39 (2002).

\bibitem{Bloch2008}
I.~Bloch, J.~Dalibard, and W.~Zwerger, \Doi{10.1103/RevModPhys.80.885}{Rev.
  Mod. Phys.} \textbf{80}, 885 (2008).

\bibitem{Fisher1989}
M.~P.~A. Fisher, P.~B. Weichman, G.~Grinstein, and D.~S. Fisher,
  \Doi{10.1103/PhysRevB.40.546}{Phys. Rev. B} \textbf{40}, 546 (1989).

\bibitem{Gurarie2009}
V.~Gurarie, L.~Pollet, N.~V. Prokof'ev, B.~V. Svistunov, and M.~Troyer,
  \Doi{10.1103/PhysRevB.80.214519}{Phys. Rev. B} \textbf{80}, 214519 (2009).

\bibitem{Sanchez-Palencia2010}
L.~Sanchez-Palencia and M.~Lewenstein, \Doi{10.1038/nphys1507}{Nat. Phys.}
  \textbf{6}, 87 (2010).

\bibitem{Stasinska2012}
J.~Stasińska, P.~Massignan, M.~Bishop, J.~Wehr, A.~Sanpera, and M.~Lewenstein,
  \Doi{10.1088/1367-2630/14/4/043043}{New J. Phys.} \textbf{14}, 043043 (2012).

\bibitem{Lee1990}
D.~K.~K. Lee and J.~M.~F. Gunn, \Doi{10.1088/0953-8984/2/38/004}{J. Phys.:
  Condens. Matter} \textbf{2}, 7753 (1990).

\bibitem{Graham2009}
R.~Graham and A.~Pelster, \Doi{10.1142/S0218127409024451}{International Journal
  of Bifurcation and Chaos (IJBC)} \textbf{19}, 2745 (2009).

\bibitem{Pilati2010}
S.~Pilati, S.~{Giorgini}, M.~{Modugno}, and N.~{Prokof'ev},
  \Doi{10.1088/1367-2630/12/7/073003}{New J. Phys.} \textbf{12}, 073003 (2010).

\bibitem{Oosten2001}
D.~van Oosten, P.~van~der Straten, and H.~T.~C. Stoof,
  \Doi{10.1103/PhysRevA.63.053601}{Phys. Rev. A} \textbf{63}, 053601 (2001).

\bibitem{Huang1992}
K.~Huang and H.-F. Meng, \Doi{10.1103/PhysRevLett.69.644}{Phys. Rev. Lett.}
  \textbf{69}, 644 (1992).

\bibitem{Giorgini1994}
S.~Giorgini, L.~Pitaevskii, and S.~Stringari,
  \Doi{10.1103/PhysRevB.49.12938}{Phys. Rev. B} \textbf{49}, 12938 (1994).

\bibitem{Kobayashi2002}
M.~Kobayashi and M.~Tsubota, \Doi{10.1103/PhysRevB.66.174516}{Phys. Rev. B}
  \textbf{66}, 174516 (2002).

\bibitem{Gaul2011_bogoliubov_long}
C.~Gaul and C.~A. M{\"u}ller, \Doi{10.1103/PhysRevA.83.063629}{Phys. Rev. A}
  \textbf{83}, 063629 (2011).

\bibitem{Hu2009}
Y.~Hu, Z.~Liang, and B.~Hu, \Doi{10.1103/PhysRevA.80.043629}{Phys. Rev. A}
  \textbf{80}, 043629 (2009).

\bibitem{White2009}
M.~White, M.~Pasienski, D.~McKay, S.~Q. Zhou, D.~Ceperley, and B.~DeMarco,
  \Doi{10.1103/PhysRevLett.102.055301}{Phys. Rev. Lett.} \textbf{102}, 055301
  (2009).

\bibitem{Pasienski2010}
M.~Pasienski, D.~McKay, M.~White, and B.~DeMarco, \Doi{10.1038/nphys1726}{Nat.
  Phys.} \textbf{6}, 677 (2010).

\bibitem{Beeler2012}
M.~C. {Beeler}, M.~E.~W. {Reed}, T.~{Hong}, and S.~L. {Rolston},
  \Doi{10.1088/1367-2630/14/7/073024}{New J. Phys.} \textbf{14}, 073024 (2012).

\bibitem{Astrakharchik2011}
G.~E. Astrakharchik and K.~V. Krutitsky, \Doi{10.1103/PhysRevA.84.031604}{Phys.
  Rev. A} \textbf{84}, 031604 (2011).

\bibitem{Ray2012}
U.~Ray and D.~M. Ceperley,
  \href{http://arxiv.org/abs/1209.1053}{arXiv:1209.1053}.

\bibitem{Muller2012_momdis}
C.~A. {M{\"u}ller} and C.~{Gaul}, \Doi{10.1088/1367-2630/14/7/075025}{New J.
  Phys.} \textbf{14}, 075025 (2012).

\bibitem{Singh1994}
K.~G. Singh and D.~S. Rokhsar, \Doi{10.1103/PhysRevB.49.9013}{Phys. Rev. B}
  \textbf{49}, 9013 (1994).

\bibitem{Gavish2005}
U.~Gavish and Y.~Castin, \Doi{10.1103/PhysRevLett.95.020401}{Phys. Rev. Lett.}
  \textbf{95}, 020401 (2005).

\bibitem{Gadway2011}
B.~Gadway, D.~Pertot, J.~Reeves, M.~Vogt, and D.~Schneble,
  \Doi{10.1103/PhysRevLett.107.145306}{Phys. Rev. Lett.} \textbf{107}, 145306
  (2011).

\bibitem{Roati2008}
G.~Roati, C.~D{'}Errico, L.~Fallani, M.~Fattori, C.~Fort, M.~Zaccanti,
  G.~Modugno, M.~Modugno, and M.~Inguscio, \Doi{10.1038/nature07071}{Nature}
  \textbf{453}, 895 (2008).

\bibitem{Modugno2009}
M.~Modugno, \Doi{10.1088/1367-2630/11/3/033023}{New J. Phys.} \textbf{11},
  033023 (2009).

\bibitem{Sanchez-Palencia2006}
L.~Sanchez-Palencia, \Doi{10.1103/PhysRevA.74.053625}{Phys. Rev. A}
  \textbf{74}, 053625 (2006).

\bibitem{Yukalov2009}
V.~I. {Yukalov}, \Doi{10.1134/S1054660X09010010}{Laser Physics} \textbf{19}, 1
  (2009).

\bibitem{Kuhn2007a}
R.~Kuhn, O.~Sigwarth, C.~Miniatura, D.~Delande, and C.~M{\"u}ller,
  \Doi{10.1088/1367-2630/9/6/161}{New J. Phys.} \textbf{9}, 161 (2007).

\bibitem{Gaul2009a}
C.~Gaul, N.~Renner, and C.~A. M\"{u}ller,
  \Doi{10.1103/PhysRevA.80.053620}{Phys. Rev. A} \textbf{80}, 053620 (2009).

\bibitem{Lugan2007a}
P.~Lugan, D.~Cl{\'e}ment, P.~Bouyer, A.~Aspect, and L.~Sanchez-Palencia,
  \Doi{10.1103/PhysRevLett.99.180402}{Phys. Rev. Lett.} \textbf{99}, 180402
  (2007).

\bibitem{Lugan2011}
P.~Lugan and L.~Sanchez-Palencia, \Doi{10.1103/PhysRevA.84.013612}{Phys. Rev.
  A} \textbf{84}, 013612 (2011).

\bibitem{Lopatin2002}
A.~V. Lopatin and V.~M. Vinokur, \Doi{10.1103/PhysRevLett.88.235503}{Phys. Rev.
  Lett.} \textbf{88}, 235503 (2002).

\bibitem{Falco2007}
G.~M. Falco, A.~Pelster, and R.~Graham, \Doi{10.1103/PhysRevA.75.063619}{Phys.
  Rev. A} \textbf{75}, 063619 (2007).

\end{thebibliography}

\end{document}